# Structural information from multilamellar liposomes at full hydration: full q-range fitting with high quality X-ray data

Georg Pabst, Michael Rappolt, Heinz Amenitsch and Peter Laggner

Institute of Biophysics and X-ray Structure Research, Austrian Academy of Sciences, Steyrergasse 17, A-8010 Graz, Austria.
(E-mail: Peter.Laggner@oeaw.ac.at).

# Abstract

We present a novel method for analyzing Small Angle X-ray Scattering data on multilamellar phospholipid bilayer systems at full hydration. The method utilizes a modified Caillé theory structure factor in combination with a Gaussian model representation of the electron density profile such that it accounts also for the diffuse scattering between Bragg peaks. Thus, the method can retrieve structural information even if only a few orders of diffraction are observed. We further introduce a new procedure to derive fundamental parameters, such as area per lipid, membrane thickness, and number of water molecules per lipid, directly from the electron density profile without the need of additional volumetric measurements. The theoretical apparatus is applied to experimental data on 1-palmitoyl-2-oleoyl-*sn*-glycero-3-phosphocholine and 1,2-dipalmitoyl-*sn*-glycero-3-phosphoethanolamine liposome preparations.

# I. INTRODUCTION

Phospholipids are the main constituents of biological membranes by forming the structural matrix into which functional membrane units such as proteins are imbedded. Among the various structures that are formed by phospholipid membranes, the lamellar liquid crystalline phase is the biologically most relevant one. The interest in the structure and physical properties of this particular phase has therefore always been an important subject in biophysical and biochemical research, since the structure is directly related to the function of the molecular aggregates. But not only the efforts to understand the function of biological membranes drive the progress in phospholipid structure research, also phospholipid-based rational drug design and bio-mimetic material development rely on physical interaction predictions.

The structural characterization of phospholipid model membranes was initiated by the pioneering work of Luzzati and coworkers [1, 2] on unoriented multilayers of diacyl-phosphocholines and was followed by a large number of X-ray and neutron scattering experiments on different phospholipid bilayer structures [3, 4]. However, the major difficulties in obtaining accurate structural data arise, apart from thermal disorder ("disorder of first kind"), from disorder in the crystal lattice ("disorder of second kind"), which is mostly dominant in the liquid crystalline phases due to their liquid properties. Two theories have been developed to model the lattice structure factor of model membranes, both accounting for the deficiencies in long range order: the paracrystalline theory (PT), a general theory for disorder of first and the second kind originated by Hosemann and Bagchi [5] and Guinier [6], and the Caillé theory (CT) [7], which was invoked for smectic liquid crystals only. The main difference between the two models is that the paracrystalline theory describes the stochastic fluctuations of the single, ideally flat layers, whereas the Caillé theory considers also bilayer undulations by applying a Hamiltonian description derived from the free energy density of a lipid bilayer, originally derived by De Gennes [8]. In 1994 the Caillé theory was modified by Zhang et al. [9] (MCT), in order to take the finite size of the lamellar stack into account; a similar expression was obtained by Nallet et al. [10]. Both theories (PT & MCT) have been applied to experimental data [10-17], but with the help of the high resolution capabilities of modern synchrotron radiation sources the superiority of the Caillé theory could clearly be demonstrated [15]. The facts therefore encourage to use MCT for smectic A liquid crystals, and moreover tests on our own data gave better fits for MCT than for a PT model (results not shown).

But having an employed theory describing well the crystal lattice and thus the position and shapes of the diffraction peaks does not overcome a principle problem of liquid crystallography: As a consequence of the lattice disorder multilamellar liposomal suspensions give hardly rise to a sufficient number of diffraction orders to derive structural information. Among the zwitterionic phospholipids the situation is somewhat better for phosphatidylethanolamine (PE) membrane stacks, exhibiting 4 Bragg peaks throughout the whole $L_\alpha$-phase, whereas the higher water content in phosphatidylcholine (PC) bilayers leads to a higher lattice disorder and thus to even less diffraction peaks observed. As a consequence, the electron density profiles are very poor in detail and likely to be affected by Fourier truncation errors. There are two ways to circumvent this problem, both applying osmotic pressure techniques. (1) One is to incubate multilamellar liposomes in aqueous solutions containing various concentrations of large, neutral polymers such as dextran or polyvinylpyrrolidone (PVP) [14-21]. With such "swelling experiments" the system is partly dehydrated, and consequently the number of observed diffraction orders increases. Structural parameters for the fully hydrated bilayer are then obtained by extrapolating the areas per lipid, derived from the partly dehydrated systems to full hydration [14-17]. (2) Even more structural information can be obtained by exposing oriented multilayers to constant relative humidity atmospheres [21-25] and depending on the degree of hydration, up to 10 diffraction orders have been recorded [23, 24]. The electron density profile from such experiments is much richer in information and even allows for a quasimolecular modeling, first applied by Wiener and White [23, 26, 27]. The phospholipid molecule is partitioned into quasimolecular fragments, and the contribution of each fragment to the bilayer profile is modeled by a Gaussian distribution. In this manner structural details have been obtained by a joint refinement of neutron and X-ray data sets [23]. Still, the major drawback of measuring oriented sample in humidity chambers is that the bilayer repeat does not swell to the value reached in the unoriented case under excess water conditions, even at 100% relative humidity. Consequently, the fully hydrated $L_\alpha$-phase cannot be exploited with this technique. The so-called "vapor pressure paradox" has been for a long time a disputed topic in the lipid community. Recently, Katsaras installed a new cell for oriented bilayers [28] and could demonstrate that the vapor pressure paradox originates simply from experimental inadequacy and has no theoretical background [29]. Hence the ghost of the vapor pressure paradox ceased haunting through the brains of lipid scientists and diffraction experiments on oriented membrane stacks will be of prime importance in future phospholipid structure research.

However, unoriented multilamellar liposomes at full hydration are still a frequent measurement situation. Not least simulations of biological systems and development of new drugs, e.g., carrier systems, will always demand the work with liposomal dispersions in the excess water situation. Here the information content is very low, if only Bragg-peaks are considered in the data analysis. We invoke a model that accounts also for the diffuse scattering of the bilayer between diffraction peaks, and thus, exploits the complete data recorded in a continuous q-range. In this way our method is capable to retrieve fundamental structural parameters, such as membrane thickness, area per lipid, and number of waters, even under above conditions, when only a few orders of diffraction are observed. We further introduce a procedure, based on simple geometric relationships, to calculate the above named parameters directly from a electron density model of the bilayer, without the need of extra volumetric measurements.

## II. THEORY

The intensity scattered from a finite stack of unoriented bilayers is described by

$$I(q) \propto \frac{\langle |f(q)|^2 s(q) \rangle}{q^2}, \quad (1)$$

where q is the absolute value of the scattering vector ($q = 4\pi \sin\theta/\lambda$), $f(q)$ the form factor and $s(q)$ the structure factor. The form factor characterizes the electron density distribution and is given in the case of a layered structure by the Fourier transform

$$f(q) = \int \rho(z) \exp(i\, q\, z) dz \quad (2)$$

of the electron density profile $\rho$ along the z-axis. The structure factor accounts for the crystalline or quasi-crystalline nature of the lattice of the bilayer stack in the liquid crystalline phase. Both, structure and form factor, are averaged over the bilayer fluctuations. By assuming that the fluctuations within the bilayer are independent of the fluctuations of the lattice points, the structure factor and the form factor can be treated separately according to Debye [6]

$$I(q) \propto \frac{1}{q^2}\left[\left|\langle f(q)\rangle\right|^2 \langle s(q)\rangle + N\left(\langle|f(q)|^2\rangle - |\langle f(q)\rangle|^2\right)\right]. \tag{3}$$

The last term in Eq. (3) gives rise to a diffuse scattering and is usually neglected, when structural information is derived from Bragg peaks only. The standard data analysis procedure is then to fit the Bragg reflections with the appropriate structure factor multiplied by a constant form factor for each single peak, which is a reasonable assumption in the vicinity of the diffraction peaks only. The electron density profile relative to the constant electron density of the buffer (water) is calculated by the Fourier synthesis

$$\rho^*(z) = \sum_{h=1}^{h\,\mathrm{max}} \pm F_h \cos\left(\frac{2\pi h z}{d}\right), \tag{4}$$

wherein $h$ is the order of reflection and $d$ the size of the unit cell.

We invoke a model that tries to solve the problem in the backward direction by means of an inverse Fourier transform. Since we record data in a continuous q-range, we should rather model the scattering function $I(q)$ in the whole range studied. The electron density profile - at a given resolution of 4 diffraction orders – can be modeled according to Wiener et al. [30] by a summation of 2 Gaussians, each representing the polar headgroup and the methyl terminus[1], respectively

$$\rho(z) = \rho_{CH_2} + \bar{\rho}_H\left[\exp\left(-\frac{(z-z_H)^2}{2\sigma_H^2}\right) + \exp\left(-\frac{(z+z_H)^2}{2\sigma_H^2}\right)\right] + \bar{\rho}_C \exp\left(-\frac{z^2}{2\sigma_C^2}\right), \tag{5}$$

where the electron densities of the headgroup $\bar{\rho}_H$ and hydrocarbon tails $\bar{\rho}_C$ are defined relative to the methylene electron density $\rho_{CH_2}$

---

[1] Wiener and White were able to model the bilayer profile with a summation of 8 Gaussians [23] representing quasi-molecular phospholipid fragments for oriented dioleoylphosphatidylcholine bilayers at 66% RH. However, this model is not applicable for the present case of the resolution limit of 4 and less diffraction orders.

$$\bar{r}_H \equiv r_H - r_{CH_2}$$

$$\bar{r}_C \equiv r_C - r_{CH_2}, \tag{6}$$

(Fig. 1). The position of the Gaussian peak is at $z_i$ ($i = H, C$; $z_C = 0$), with a standard deviation of $\sigma_i$. The form factor of this electron density model can be calculated analytically by applying Eq. (2)

$$\langle f(q) \rangle = F(q) = 2F_H(q) + F_C(q), \tag{7}$$

where the individual components denote the form factor of the headgroup

$$F_H(q) = \sqrt{2\pi} \sigma_H \bar{r}_H \exp\left(-\frac{\sigma_H^2 q^2}{2}\right) \cos(qz_H) \tag{8}$$

and the form factor of the hydrocarbon chains

$$F_C(q) = \sqrt{2\pi} \sigma_C \bar{r}_C \exp\left(-\frac{\sigma_C^2 q^2}{2}\right). \tag{9}$$

Equation (7) gives the time averaged form factor of the bilayer as a continuous function of the scattering vector $q$.

Since the structure factor retained from the Caillé theory considers the lattice disorder, a full q-range description will also account for the diffuse scattering term in Eq. (3). We choose the discrete formula of the MCT structure factor [9] in the equivalent form of

$$\langle s(q) \rangle = S(q) = N + 2\sum_{k=1}^{N-1}(N-k)\cos(kqd)e^{-\left(\frac{d}{2\pi}\right)^2 q^2 \eta_1 \gamma}(\pi k)^{-\left(\frac{d}{2\pi}\right)^2 q^2 \eta_1}, \tag{10}$$

given in a manuscript of Lemmich et al. [31]. The mean number of coherent scattering bilayers in the stack is denoted as $N$, $\gamma$ is Eulers constant. The Caillé parameter $\eta_1$ involves both, the bending modulus $K$ of lipid bilayers and the bulk modulus $B$ for compression [7, 9]

$$\eta_1 = \frac{q^2 kT}{8\pi \sqrt{KB}} \tag{11}$$

with

$$\mathbf{h}_h = \mathbf{h}_1 h^2. \tag{12}$$

However, we discovered during our data analysis an additional diffuse scattering contribution, which is not described by the MCT. Its origin is attributed to bilayers with strong lattice defects or unilamellar vesicles, which display neither short-range nor (quasi) long-range order. The total scattered intensity is therefore given by the diffraction of the phospholipid multilayers within the quasi long-range order lattice, plus the additional diffuse scattering of single, uncorrelated bilayers

$$I(q) \propto \frac{1}{q^2}\left(|F(q)|^2 S(q) + N_{diff}|F(q)|^2\right). \tag{13}$$

We will refer in the further context of this article to the above described model as MCG, since it is a combination of MCT and a Gaussian electron density representation of the headgroup [30].

A further benefit of this method is that one can derive structural parameters from simple geometric relationships, without the need of volumetric data as, e.g., in the approach of McIntosh and Simon [32], or Nagle et al. [14]. For determining the area per lipid, we follow the formalism given by Lemmich et al. [33] by calculating the ratio $\tilde{r}_r \equiv \overline{r}_H / \overline{r}_C$ (see Eq. (5)), which yields

$$A = \frac{1}{\mathbf{r}_{CH_2}(\tilde{r}_r - 1)}\left(\frac{\tilde{r}_r n_C^e}{d_C} - \frac{n_H^e}{d_H}\right), \tag{14}$$

where $n_C^e$ is the number of hydrocarbon electrons and $n_H^e$ the number of headgroup electrons, respectively. The headgroup size $d_H$ can be estimated from the FWHM of the Gaussian, representing the headgroup and the hydrocarbon chain length $d_C$ can be derived from

$$d_C = z_H - \frac{FWHM_H}{2}. \tag{15}$$

Further parameters of interest are the bilayer thickness

$$d_B = 2\left(z_H + \frac{FWHM_H}{2}\right), \tag{16}$$

the thickness of the water layer

$$d_W = d - d_B, \tag{17}$$

and the number of interbilayer free water per lipid molecule

$$n_W^* = \frac{A d_W}{2 V_W} \tag{18}$$

(see, e.g., [1, 14, 32]), where $V_W$ is the volume of one water molecule (approx. 30 Å$^3$). The total number of water, including the molecules intercalated into the bilayer, can be estimated from the distance of the headgroup to the bilayer center $z_H$

$$n_W = \frac{A(d/2 - z_H)}{V_W}. \tag{19}$$

Finally, the electron density profile can be set on an absolute scale. Here we follow the procedure introduced by Nagle and Wiener [34] by calculating the integral

$$A\int_0^{d/2}(r(z) - r_{CH_2})dz = A\alpha \int_0^{d/2}\left[\bar{r}_H \exp\left(-\frac{(z-z_H)^2}{2s_H^2}\right) + \bar{r}_C \exp\left(-\frac{z^2}{2s_C^2}\right)\right]dz, \tag{20}$$

wherein $\alpha$ is the instrumental scaling constant. The evaluation of the left integral gives

$$A\int_0^{d/2}(r(z) - r_{CH_2})dz = n_L^e + n_W^e - \frac{r_{CH_2} A d}{2}, \tag{21}$$

with $n_L^e$ being the number of electrons of the phospholipid molecule and $n_W^e$ the number of water electrons, i.e., the total number of waters per lipid molecule times the number of electrons in one water molecule. The Integral on the right is given by

$$\Gamma = \int_0^{d/2}\left[\bar{r}_H \exp\left(-\frac{(z-z_H)^2}{2s_H^2}\right) + \bar{r}_C \exp\left(-\frac{z^2}{2s_C^2}\right)\right]dz$$

$$= \sqrt{\frac{p}{2}} \bar{r}_H s_H \left[ \mathrm{erf}\left(\frac{d/2 - z_H}{\sqrt{2} s_H}\right) - 2\,\mathrm{erf}\left(\frac{a - z_H}{\sqrt{2} s_H}\right) - \mathrm{erf}\left(\frac{z_H}{\sqrt{2} s_H}\right) \right]$$

$$+ \sqrt{\frac{p}{2}} \bar{r}_C s_C \left[ \mathrm{erf}\left(\frac{d/2}{\sqrt{2} s_C}\right) - 2\,\mathrm{erf}\left(\frac{a}{\sqrt{2} s_C}\right) \right], \tag{22}$$

the parameter $a$ is the root of the function $r(z) - r_{CH_2}$. By combining both results, Eq. (21) and (22), one arrives at

$$a = \frac{n_L^e + n_W^e - \dfrac{r_{CH_2} A d}{2}}{A \Gamma} \tag{23}$$

for the instrumental scaling constant. The electron density on an absolute scale is then given by

$$r_{abs}(z) = r_{CH_2} + a \left\{ \bar{r}_H \left[ \exp\left(-\frac{(z - z_H)^2}{2 s_H^2}\right) + \exp\left(-\frac{(z + z_H)^2}{2 s_H^2}\right) \right] + \bar{r}_C \exp\left(-\frac{z^2}{2 s_C^2}\right) \right\} \tag{24}$$

(cf. Eq. (5)).

## III. EXPERIMENTAL METHODS

### A. Sample preparation

1-palmitoyl-2-oleoyl-*sn*-glycero-3-phosphocholine (POPC) and 1,2-dipalmitoyl-*sn*-glycero-3-phosphoethanolamine (DPPE) were purchased from Avanti Polar Lipids, Birmingham Alabama, and used without further purification. Multilamellar liposomes were prepared by dispersing weighted amounts of dry lipids, typically 20-30% w/w, in bidistilled water. To ensure complete hydration, the lipid dispersions were incubated for about 4 hours at least 10 °C above the main transition temperature. During this period the lipid dispersions were vigorously vortexed. Aqueous dispersions of this lipids display narrow, cooperative melting transitions within the limits of published values, thus proving that the lipid purity corresponds to the claimed one of 99%. The POPC dispersions were further subjected to a centrifugation (centrifuge: 3K18, Sigma, Germany / rotor: 12 x 1.5 (max. 2.2 ml) / time: 10 min / 12000 rpm) to determine the content of unilamellar vesicles [35]. The phospholipid content in the supernatant was assayed by an enzymatic kit test (Phospholipides enzymatiques PAP 150, bioMérieux, France). A proportion of 0.1-0.2% of the total phospholipids was found as unilamellar vesicles in the supernatant. Thus, diffuse scattering from unilamellar vesicles can be neglected.

### B. Experimental protocol

Small angle X-ray scattering (SAXS) experiments were carried out at the SAXS beam-line, ELETTRA [36, 37]. The diffraction patterns were recorded with a one-dimensional position sensitive detector [38] monitoring the q-range between $2\pi/90$ and $2\pi/10$ Å$^{-1}$ at a photon energy of 8 keV. The lipid dispersions were kept in a thin-walled 1 mm diameter Mark capillary held in a steel cuvette, which provides good thermal contact to the Peltier heating unit. Exposure times were typically in the range of 5 minutes. Random thin layer chromatography tests for radiation damage resulted normal, i.e., they showed no decomposition products. The position calibration of the detector was performed by using the diffraction pattern of silver behenate powder ($CH_3(CH_2)_{20}COOAg$) (repeat unit = 58.38 Å) [39].

### C. Data analysis

The X-ray data was analyzed in terms of the model developed in section II. After substracting the background scattering from water and the sample cell, we applied the following procedure. First, the Bragg reflections were fitted by Lorentzians taking the square root of the peak area as an estimate for the constant form factor of each peak. Utilizing Eq. (4) a raw electron density profile was calculated with the appropriate phases (- - + - -) [24, 32]. The profile was then fitted with the electron density model Eq. (5), taking the results as input parameters for the further calculations. Thereafter, the diffraction pattern was fitted in the complete q-range by operating Eq. (7) and (10), where the finite instrumental resolution has to be accounted for by the convolution

$$I_{obs}(q) = b \int_{-\infty}^{+\infty} I(q') r(q - q') dq', \qquad (25)$$

$b$ is the instrumental scaling constant. We chose an instrumental resolution function $r$ with a Gaussian profile

$$r(q) = \exp\left(-\frac{q^2}{2s_r^2}\right), \qquad (26)$$

where the standard deviation $s_r$ is typically in the range of $1.2 \cdot 10^{-3}$ Å$^{-1}$ for the given experimental set-up. The number of fit parameters is 9 compared to 8 for the MCT model at 4 orders of diffraction [9]. Least square fitting was performed with self-written IDL (Interactive Data Language) procedures, utilizing MPFIT [40], which is based on the MINPACK library [41]. Structural parameters have been calculated according to Eqs. (14) – (19).

## IV. EXPERIMENTAL RESULTS

We measured X-ray diffraction profiles from unoriented liposomal suspensions of POPC and DPPE at 20 and 30% w/w lipid concentration, respectively. Both phospholipid samples were measured in the lamellar liquid crystalline phase (smectic A); POPC was equilibrated at 2°C and 50°C, DPPE at 75°C, respectively.

Figure 2 shows the diffraction pattern of POPC. Diffraction orders number 1, 2, 3, and 5 are observed, the 4$^{th}$ order is ruled out by the form factor. The background between the Bragg reflections is clearly modulated by the bilayer form factor, most dominantly between the first and third order. The solid line gives the best fit of the MCG model, developed in the theory section (Eqs. (1), (7), (10), and (25)). The results for the fit parameters are given in the second column of Table I. Note, that no diffuse background is fitted. The system has been equilibrated at 2°C, only, and hence lattice defects are much more suppressed than at higher temperatures, where molecular motions are more destructive to the lattice order. Figure 2 depicts further the MCT fit (dashed line) within a q-range of ± 0.01 Å$^{-1}$ around each Bragg peak (cf. [14]); a close view of the first-order peak is drawn in the insert to Fig. 2. The comparison demonstrates two facts: First, standard MCT uses only a small fraction of the available diffraction data. Second, MCT gives a better fit for the peak tops, but a poorer fit for the peak tails, as it applies a constant form factor within the fitted peak region. Neither of the model functions perfectly describes the experimental data points. With the MCT method it is apparently easier to model the scattered intensity in a limited regime around the Bragg peaks, while MCG proved to be better suited to model the asymmetric tails. A quantitative comparison of the two models in terms of the respective, reduced $\chi^2$ sums is not expedient, as different numbers of data points are being considered. It is more important to state that MCG gives a qualitatively good fit for the full q-range, i.e., the diffraction peaks including the diffuse scattering, whereas MCT works in the vicinity of Bragg peaks only.

Figure 3 shows the differences between MCT and MCG in terms of the electron density profiles. The Fourier synthesis for the MCT fit shows an anomalous, small hump at the center of the water layer, due to truncation errors. The MCG model, on the other hand, gives a smoother representation of the bilayer profile, since it excludes by definition Fourier truncation errors (Eq. (5)). However, with 4 diffraction orders given, both profiles yield similar structure results. Thus full advantage of MCG can be taken only on data with less Bragg peaks.

At 50°C the scattered intensity of POPC exhibits different features (Fig. 4). Evidently, the number of clearly recognizable diffraction orders has decreased from 4 to 2, an effect which is attributed to stronger thermal induced fluctuations of the bilayers, but not only. The position of the 3$^{rd}$ order Bragg peak is close to a minimum of the bilayer form factor, therefore the 3$^{rd}$ order is also attenuated because of the bilayer structure. Applying Fourier methods, such as MCT, gives in this case only very rough structural information, as only 2 diffraction orders can be used to construct the electron density profile (cf. insert to Fig. 4, dashed line). The MCG model (solid line), on the other hand, gives a clearly refined picture of the bilayer, which affects especially the headgroup region, whereas the terminating methylene group remains strongly smeared. Further, one should expect a diffuse scattering from lattice defects, as the temperature has increased from 2°C to 50°C. Indeed, we find a diffuse contribution of the bilayer form factor (cf. Table I). An additional fingerprint for enhanced fluctuations at higher temperatures is the Caillé parameter $\boldsymbol{h_1}$, which is almost 2 times greater than at 2°C.

Compared to POPC, the diffraction pattern of DPPE (Fig. 5) exhibits a completely different characteristic, regarding both, the number of observed Bragg peaks - here we detect the first 4 orders - as well as the diffuse background between the reflections. The solid line gives again the best fit of the MCG model. The fit is in good agreement with the experimental data, the fit results are given in Table I. The model fits also here a contribution of diffuse scattering, which is again attributed to the enhanced molecular motions at 75°C. The insert to Fig. 5 illustrates the effect of the MCG on Fourier artifacts. The unreal Fourier ripples of the Lorentzian model (dashed line), a consequence of the Fourier synthesis with 4 terms only, are suppressed resulting in a smooth bilayer profile (solid line) that corresponds to the resolution of the experiment.

Further structural parameters have been calculated according to the geometric considerations expressed in Eqs. (14)-(19). The number of headgroup electrons is 164 and the number of hydrocarbon chain electrons is 256 for POPC, whereas $n_H^e = 140$ and $n_C^e = 242$ for DPPE, respectively. The methylene electron density is $0.317 \pm 0.003$ e/Å$^3$ according to Wiener et al. [30]. The results for the two measured samples are listed in Tab. II. The structural parameters of POPC at 2°C are compared to the values obtained by the volumetric method, which was introduced by McIntosh and Simon [32, 42] for phospatidylethanolamines and further adopted for lecithins by Nagle et al. [14]. A brief description of the formalism is given in the Appendix. For the lipid volume, which is an input parameter of the method, we refer to the measurement of Hianik et al. [43] and extrapolate to 2°C, so that we get $V_L^l = 1223$ Å$^3$. Within measurement errors, which are larger for the volumetric method, mostly due to uncertainties in the headgroup thickness [12, 13] both methods result in the same values for the structural parameters (cf. column 1 & 2 of Tab. II). At 50°C, the repeat distance is reduced by 2 Å, and the bilayer thickness by approx. 8 Å. On the other hand, the interbilayer water thickness is increased by roughly 6 Å, a sign for water uptake from the excess phase as observed in the increase of parameter $n_W$ or $n_W^*$, respectively, due to reduced van der Walls interactions between opposing bilayers [44] at stronger undulations [45]. A further parameter, which increases with temperature is the area per lipid. The structural results for DPPE, give a very thin water layer of 10 water molecules per lipid molecule out of which approx. 6 are intercalated into the bilayer. These values are in good agreement with the data published by McIntosh and Simon for dilauroylphosphatidylethanolamine (DLPE) [32]. The small fluid space in PE bilayers could arise from interbilayer hydrogen bond formation through the water molecules or electrostatic interactions between the amine and phosphate groups of opposing bilayers [32].

Finally, the electron density profiles were put on an absolute scale by applying Eqs. (20)-(24). An input parameter is the total number of electrons per lipid molecule, which is 420 for POPC and 382 for DPPE, respectively. The results are plotted in Fig. 6; Fig. 6 (a) and 6 (b) give the absolute electron density of POPC at 2°C and 50°C, respectively, whereas Fig. 6 (c) depicts the absolute electron density of DPPE at 75°C.

## V. DISCUSSION

A new model has been introduced to analyze small angle diffraction data of unoriented phospholipid membrane stacks at high instrumental resolution. The formalism combines a form factor, related to a Gaussian representation of the electron density profile (Fig.1), with a MCT structure factor. The proposed electron density model gives the mean structure of a phospholipid bilayer time averaged over all fluctuations and is well suited to represent the X-ray picture one sees from not more than 5 orders of diffraction. Higher orders - which can be obtained by aligning the layers only - would result in a more detailed electron density profile for which other electron density model, like ,e.g., hybrid types of Gaussians and strip-models [34] would give a better representation. Such models have also been tried out on our data, but were found to fail because of too many correlating fit parameters for the given instrumental resolution. It is reasonable to model the electron density profile by means of analytic functions, as the features of its structure are well known since the pioneering work of Luzzati and Tardieu [1, 2]. The difference in the distinct phospholipid bilayer structures are then accounted for by adjusting the parameters, i.e., headgroup position, headgroup width, etc., of the analytical function. The inverse Fourier method, which takes the form factor of the bilayer model and fits it together with a structure factor to the scattered intensity has further the advantage of excluding Fourier truncation errors. The MCG model has been tested experimentally on POPC and DPPE multilayers giving good fit results [see results section & Fig. 2, Fig. 4, Fig. 5, Tab. I].

Several other models have already been published [5, 6, 9, 10, 33], in order to perform the same task. We shall briefly discuss the most prominent ones. In 1994, Nagle and coworkers introduced the modified Caillé theory and gave an experimental proof of its superiority to the classical paracrystalline theory [9, 15]. The group usually records high-resolution data at a synchrotron beam-line by means of a diffractometer, but in the vicinity of the Bragg reflections only. Electron density profiles are computed by applying the standard Fourier synthesis (Eq. (4)). In contrary, we use an equivalently brilliant source, but a detecting system, which is able to monitor the diffraction pattern in a continuous range of scattering angles. In this case, applying the standard MCT-data analysis, which works only in the regime close to diffraction peaks, means to reject all the information hidden in the diffuse background scattering between the Bragg peaks (Fig. 2). This information becomes even more valuable if less than 4 orders are observed. Nagle and coworkers report only two diffraction orders for unoriented dipalmitoylphosphatidylcholine (DPPC), egg phosphatidylcholine (EPC), dimyristolphosphatidylcholine (DMPC), and dioleoyl-phosphatidylcholine (DOPC) bilayers in excess water [14 - 17], which is insufficient to obtain satisfactory structural information, if only the Bragg peaks are considered. The common circumvention of this problem are osmotic stress experiments [14-21, 23, 24], where the system is partly dehydrated, and thus more diffraction orders are detected as bilayers are consequently hindered in undulation. Structural information of the fully hydrated phase is accessible then only through a numerical extrapolation to zero osmotic pressure. It is well known that extrapolations are always inherent to large uncertainties and should be avoided if possible. The MCG model, on the other hand, describes also the diffuse scattering and is thus capable of obtaining structural information even at low Bragg reflection information content, e.g., POPC at 50°C (Fig. 4). Moreover, the assumption of a constant form factor for each Bragg peak is not very accurate for higher diffracting orders, as peaks broaden strongly and more and more scattered intensity is smeared to the peak tails. For instance, the third order peak of the 2°C-POPC diffraction pattern displays an asymmetric shape (Fig. 2), which is obviously due to the modulation by a non-constant bilayer form factor. Such effects are not seen in the X-ray data published by Nagle and coworkers, because the observation of asymmetric peak shapes is likely to depend on the lipid type and on its specific form factor, e.g., the diffraction pattern of DPPE does not exhibit any asymmetric peaks (Fig. 4). Further, data treated with MCT only, has not always been presented in a uniform fashion, i.e., with increasing order (h = 1 to 3) decreases the data point density [16, 17] or the selected q-range [15]. Thus, peak asymmetries, even if present are difficult to be seen.

Nallet et al. [10] suggested a model similar to MCT [9] to analyze small angle scattering data on bis 2-ethylhexyl sodium sulphosuccinate (AOT) and didodecyl dimethyl ammonium bromide (DDAB) / water systems. They combined the structure factor with the form factor of a strip model for a continuous q-range fit function. Although the strip model for the AOT/water and DDAB/water systems differs somewhat from a reasonable strip model for phospholipid bilayers, this method could in principle easily be adopted with the advantage of less fit parameters. Still we refer to the common criticism on strip models, which is that discontinuous boundaries between the different regions of the bilayers are an unrealistic picture of a fluctuating bilayer.

A quite different approach was introduced by Lemmich et al. [33] for neutron scattering experiments. He proposed a strip model for the bilayer, but averaged its form factor together with a paracrystalline structure factor without decoupling the two entities as the two other theories do (Eqs. (1) and (3)). Lemmich analyzed his data in terms of both, his model and MCT, but the fits gave equally good results for phospholipids in the lamellar liquid crystalline phase. The most convincing explanation is that the strong instrumental smearing, inherent to neutron scattering experiments, does not allow for any decision. Since not even Lemmich could show better fit results for phospholipids in the $L_\alpha$-phase, we see no argument to apply his model which would imply a recalculation of the whole formalism, since X-rays "see" a different contrast than neutrons do.

Concluding the last paragraph, we should state that the models that have been discussed are without any doubts appropriate for the measurement methods applied by the individual groups. This is clearly demonstrated by the good fits to their experimental data. However, for the given reasons our method is best tailored to extract as much information as possible from high resolution X-ray data recorded in a continuous range.

A further benefit of MCG is that structural parameters like bilayer thickness, area per lipid, water distribution, etc., can be estimated from simple geometric considerations. Despite the gravimetric method of Luzzati [1], the commonly used method, initiated by McIntosh and Simon [32, 42] and applied by Nagle et al. [14], relies on additional information of the lipid volume, which is supplied by specific volumetric measurements. The algorithm is build up upon a comparison with a known gel phase structure, assuming that the volume of the headgroup is the same for both phases (cf. Appendix, Eq. (A1), (A2)). For phospholipids with a PC headgroup one usually employs the structural data of DPPC in the $L_{\beta'}$-phase, published by Sun et. al [46, 47]. A further structural input, i.e., the headgroup thickness, is needed to calculate the bilayer thickness according to the steric definition [42] (Eq. (A4)). McIntosh and Simon suggested a value of 10 Å for PC headgroups and 8 Å for PE's, derived from space filling molecular models. The headgroup conformation of DPPC has been measured by Büldt et al. [48, 49], by means of neutron diffraction and deuteron labels, but at very low water content (10 & 25 % w/w). From the published data the heagroup thickness can be extracted as $d_H = 9 \pm 1.2$ Å, a value which is employed by Nagle and coworkers, without considering the measurement error within which the values given by McIntosh and Büldt are equal. However, the headgroup conformation is likely to depend on temperature, pressure, chain tilt [30] or hydration [24], which directly affects the headgroup dimensions, so that the volume of the PC headgroup in the $L_{\beta'}$-phase is not evidently the same as in the $L_{\alpha}$-phase. Hence, a method which utilizes the assumption of constant headgroup volume and size, respectively, and even relies on measurements on systems different from the situation of fully hydrated bilayers, can be justifiable but certainly leads to a rough estimate. A way out of this dilemma should be structural data from highly aligned multilayers at full hydration according to the method of Katsaras et al. [28]. However, it is possible to obtain also reasonable estimates for unoriented systems without the need of extra data input by the simple geometric relationships of the Gaussian electron density model (Eqs. (14)-(19)). The results compare well to those obtained by the volumetric method (cf. Table II) and even display smaller errors.

The Gaussian electron density profile can be set on an absolute scale, which is often desirable. The scaling factor is computed by integrating the profile from the center to the border of the unit cell (Eq. (20)). This can be easily done, since the electron density profile is given as an analytic function. However, we argue to take absolute electron densities with great care, since the relative error of the scaling factor is large (0.2 for POPC at 50°C), a consequence of the large number of error contributors in the calculation procedure. This implies also to absolute electron densities published by other groups [14-17, 30], but has not been discussed there.

In conclusion, we remark that the MCG model gives considerable more structural information than standard MCT, provided that the number of recorded diffraction orders is less than 4. At 4 orders of diffraction one obtains equally good results (Fig. 3). The advantages of the model are due to a cancellation of Fourier artifacts, and a simple method to derive structural parameters. Since the model can retrieve structural information from the diffuse scattering its potential increases in importance, when less than four orders of diffraction are recorded (Fig. 4). This is a common situation for fully hydrated phosphaditylcholine bilayers, which include about 3 times more interbilayer water than phosphaditylethanolamine bilayer systems.

## ACKNOWLEDGEMENTS

The authors are grateful to J. F. Nagle and H. I. Petrache, for helpful discussions and for providing us the source code of the program MCT. We also express our thanks to F. Nallet and J. Lemmich for sending additional manuscripts and to H. Sormann for helpful discussions on statistics. This work has been supported by the "Elettra-Project" of the Austrian Academy of Sciences. M. Rappolt is the recipient of a long-term grant from the European Commission under the program "Training and Mobility of Researchers" [Contract no. SMT4-CT97-9024(DG12-CZJU)].

APPENDIX

Structural parameters for bilayers in the lamellar liquid crystalline phase can be derived upon the assumption that the volume of the phospholipid headgroup is equal to the volume in the gel phase [14]

$$V_H^l = V_H^g, \tag{A1}$$

where the superscript $l$ denotes the liquid phase and $g$ the gel phase. By calculating the difference in the total lipid volume $V_L^l - V_L^g$ one arrives at

$$A^l = \frac{V_L^l - V_H^g}{d_C^g + \frac{d_{HH}^l - d_{HH}^g}{2}} \tag{A2}$$

for the area of the fluid bilayer, where $d_C$ is the hydrocarbon chain length and $d_{HH}$ the head-to-head-group distance over the bilayer. For phospholipids with a PC headgroup one usually employs the structural data of $L_{\beta'}$-DPPC as published by Sun et al. [46]: $V_H^g = 319 \pm 6 \text{Å}$, $d_C^g = 17.3 \pm 0.2 \text{Å}$, and the corrected value of the head-to-head-group distance [47] $d_{HH}^g = 42.8 \pm 0.2 \text{ Å}$. The hydrocarbon chainlength is given by

$$d_C^l = \frac{V_L^l - V_H^g}{A^l} \tag{A3}$$

and the bilayer thickness, according to the steric definition of McIntosh and Simon [42], by

$$d_B^l = 2(d_C^l + d_H). \tag{A4}$$

The headgroup thickness $d_H$ has been estimated from space filling models to be 10 Å for PC's and 8 Å for PE's, whereas Büldt et al. found a value of $9 \pm 1.2$ Å with neutron diffraction experiments at a hydration of 10% w/w [48, 49]. The interbilayer water thickness and the number of free water is given according to Eqs. (17) and (18).

Sometimes it is desirable to compare the structural results with already published data derived by applying the gravimetric method of Luzzati [1]. The Luzzati bilayer thickness is calculated as

$$d_B^{Luzzati} = \frac{2V_L}{A},  \quad (A5)$$

with the corresponding interbilayer water thickness the total number of water molecules per lipid are obtained according to Eq. (19).

TABLE I. Fit results for the diffraction patterns of POPC at 2°C and 50°C, and DPPE at 75°C (cf. Fig. 1). The parameters $\bar{r}_H$ and $\bar{r}_C$ are given in absolute units according to Eq. (24) (see also Fig. 6).

| Fit parameter | POPC | | DPPE |
|---|---|---|---|
| | T = 2°C | T = 50°C | T = 75°C |
| $z_H$ (Å) | 20.2 ± 0.1 | 17.0 ± 0.3 | 19.2 ± 0.1 |
| $s_H$ (Å) | 3.6 ± 0.1 | 3.6 ± 0.2 | 3.3 ± 0.1 |
| $\bar{r}_H$ (e/Å³) | 0.11 ± 0.01 | 0.11 ± 0.01 | 0.15 ± 0.01 |
| $s_C$ (Å) | 4.8 ± 0.2 | 6.8 ± 0.7 | 2.5 ± 0.2 |
| $\bar{r}_C$ (e/Å³) | -0.08 ± 0.01 | -0.10 ± 0.02 | -0.06 ± 0.01 |
| $d$ (Å) | 66.2 ± 0.1 | 64.3 ± 0.1 | 51.4 ± 0.1 |
| $h_1$ (Å) | 0.0504 ± 0.0005 | 0.092 ± 0.001 | 0.016 ± 0.001 |
| $N$ | 28.0 ± 1.0 | 23.0 ± 1.0 | 52 ± 1 |
| $N_{diff}$ | 0.0 | 0.17 ± 0.09 | 1.08 ± 0.04 |

TABLE II. Derived structural parameters calculated by using Eqs. (14)-(19). The results for POPC at 2°C are compared to the values obtained by using the volumetric method [16, 17, 32] (cf. Appendix).

| parameter | POPC | | | DPPE |
|---|---|---|---|---|
| | T = 2°C | | T = 50°C | T = 75°C |
| | volumetric | geometric | geometric | geometric |
| $d$ (Å) | 66.2 ± 0.1 | 66.2 ± 0.1 | 64.3 ± 0.1 | 51.4 ± 0.1 |
| $d_B$ (Å) | 50.2 ± 3.6 | 48.9 ± 0.3 | 42.5 ± 1.1 | 46.2 ± 0.4 |
| $d_W$ (Å) | 16.0 ± 3.7 | 17.3 ± 0.4 | 21.7 ± 1.2 | 5.3 ± 0.5 |
| $d_C$ (Å) | 16.1 ± 0.6 | 16.0 ± 0.2 | 12.8 ± 0.6 | 15.4 ± 0.2 |
| $A$ (Å$^2$) | 56 ± 2 | 54 ± 1 | 62 ± 1 | 52 ± 1 |
| $n_W$ | 22 ± 2 | 24 ± 1 | 31 ± 1 | 11.3 ± 0.3 |
| $n_W^*$ | 15 ± 4 | 16 ± 1 | 23 ± 2 | 4.6 ± 0.4 |

Figure Captions

FIG. 1. The Gaussian electron density profile representation of a phospholipid bilayer corresponding to a X-ray resolution of 4 Bragg peaks.

FIG. 2. The best fit of the MCG model (solid line) and MCT (dashed line within marked peak region) to the diffraction pattern of POPC at 2°C. The insert gives a zoom of the first order Bragg peak.

FIG. 3. Comparison of the electron density profile for POPC bilayers at 2°C obtained by a Fourier synthesis (dashed line), using MCT and the MCG refined profile (solid line).

FIG. 4. The best fit of the MCG model (solid line) to the diffraction pattern of POPC at 50°C. The insert gives the electron density profile obtained by a Fourier synthesis (dashed line), using Lorentzians to fit the Bragg peaks, and the profile refined with MCG (solid line).

FIG. 5. The best fit of the MCG model (solid line) to the diffraction pattern of DPPE at 75°C. The insert gives the electron density profile obtained by a Fourier synthesis (dashed line), using Lorentzians to fit the Bragg peaks, and the profile refined with MCG (solid line).

FIG. 6. Absolute electron density profiles of POPC at 2°C (a), POPC at 50°C (b), and DPPE at 75°C (c). Deviations due to the error of the instrumental scaling factor *a* are depicted as a gray area enveloped by the maximal positive (dashed line) and negative (dot-dashed line) divergence.

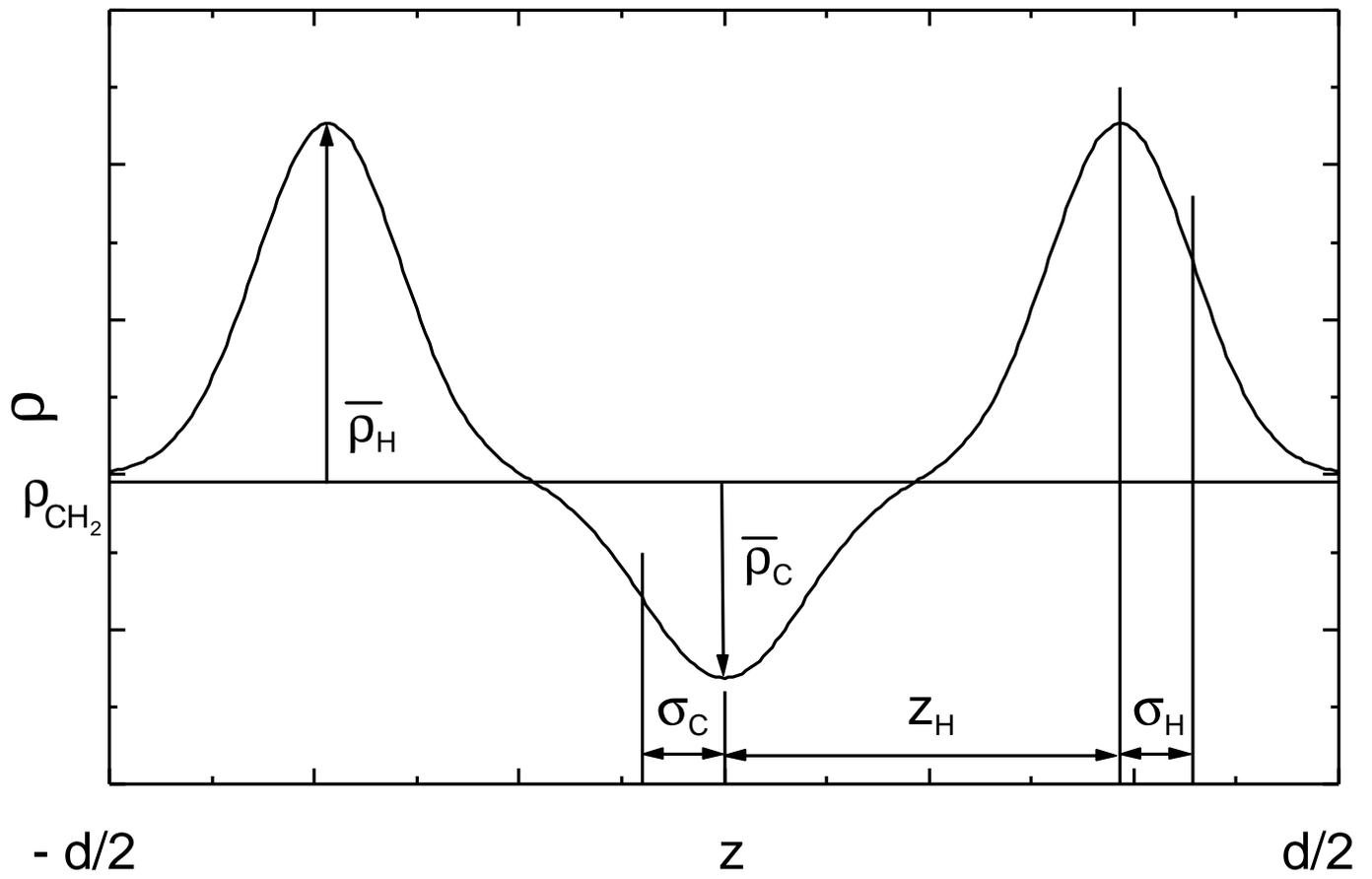

**FIG. 1** / Georg Pabst

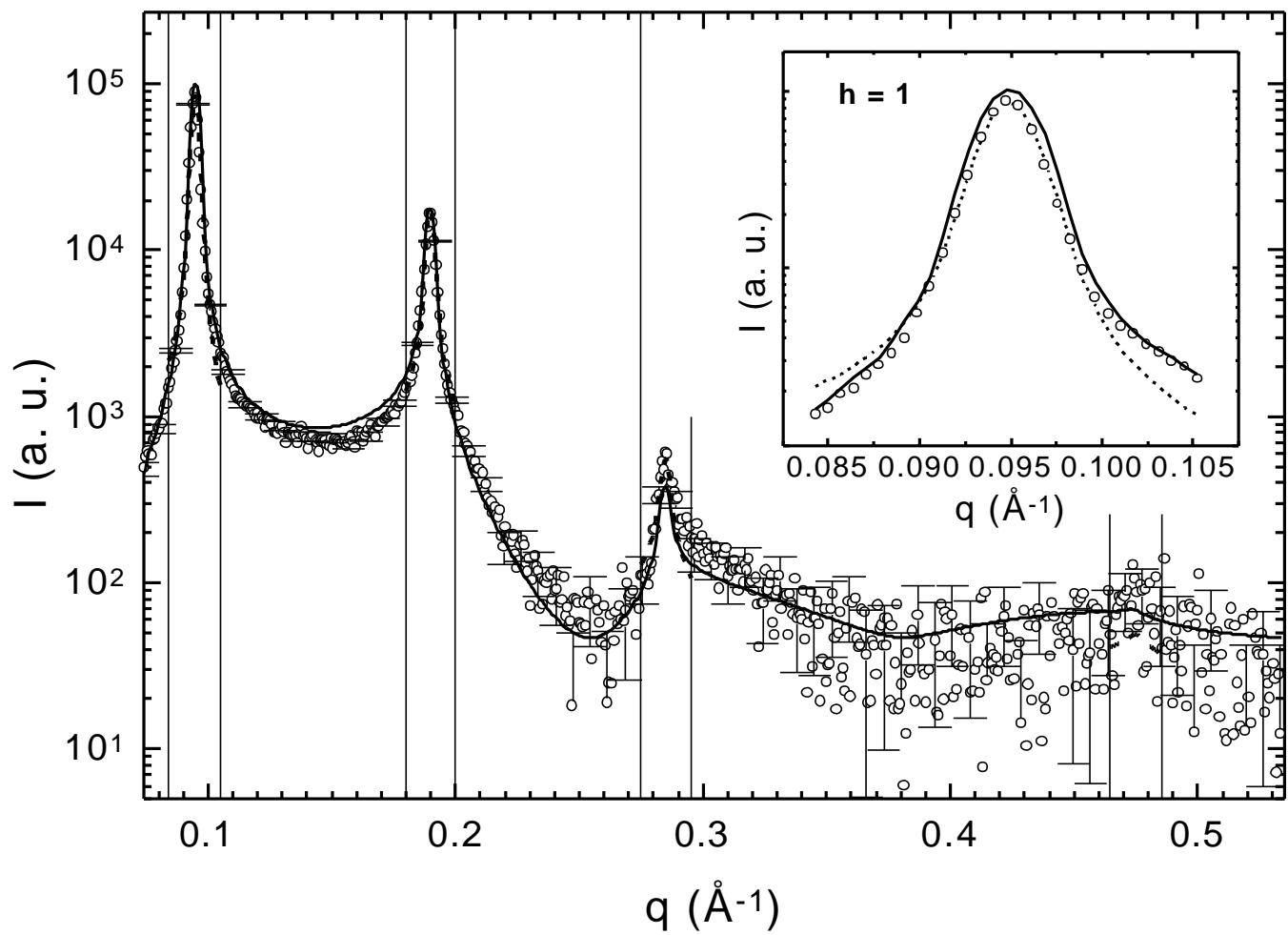

**FIG. 2** / Georg Pabst

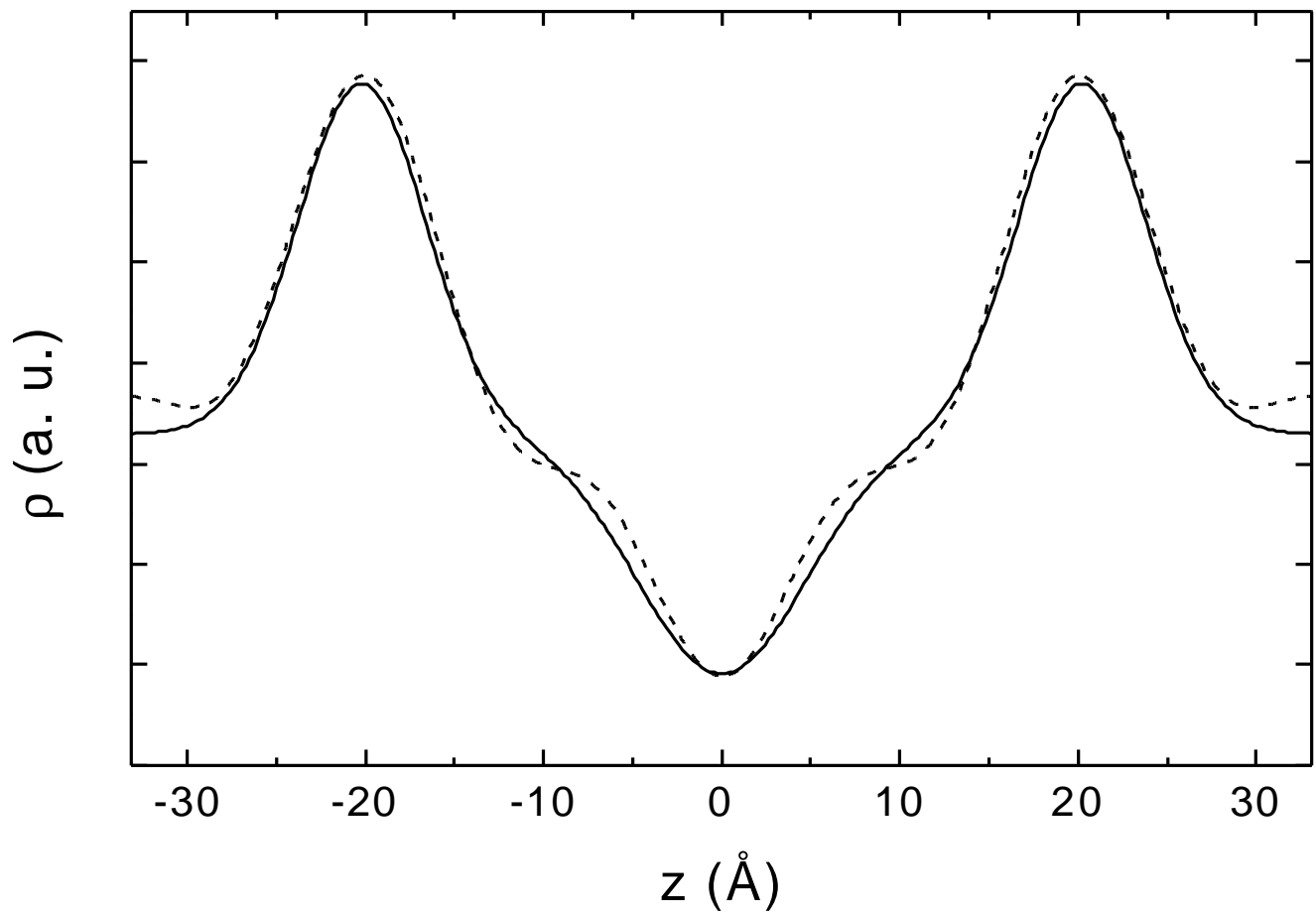

**FIG. 3** / Georg Pabst

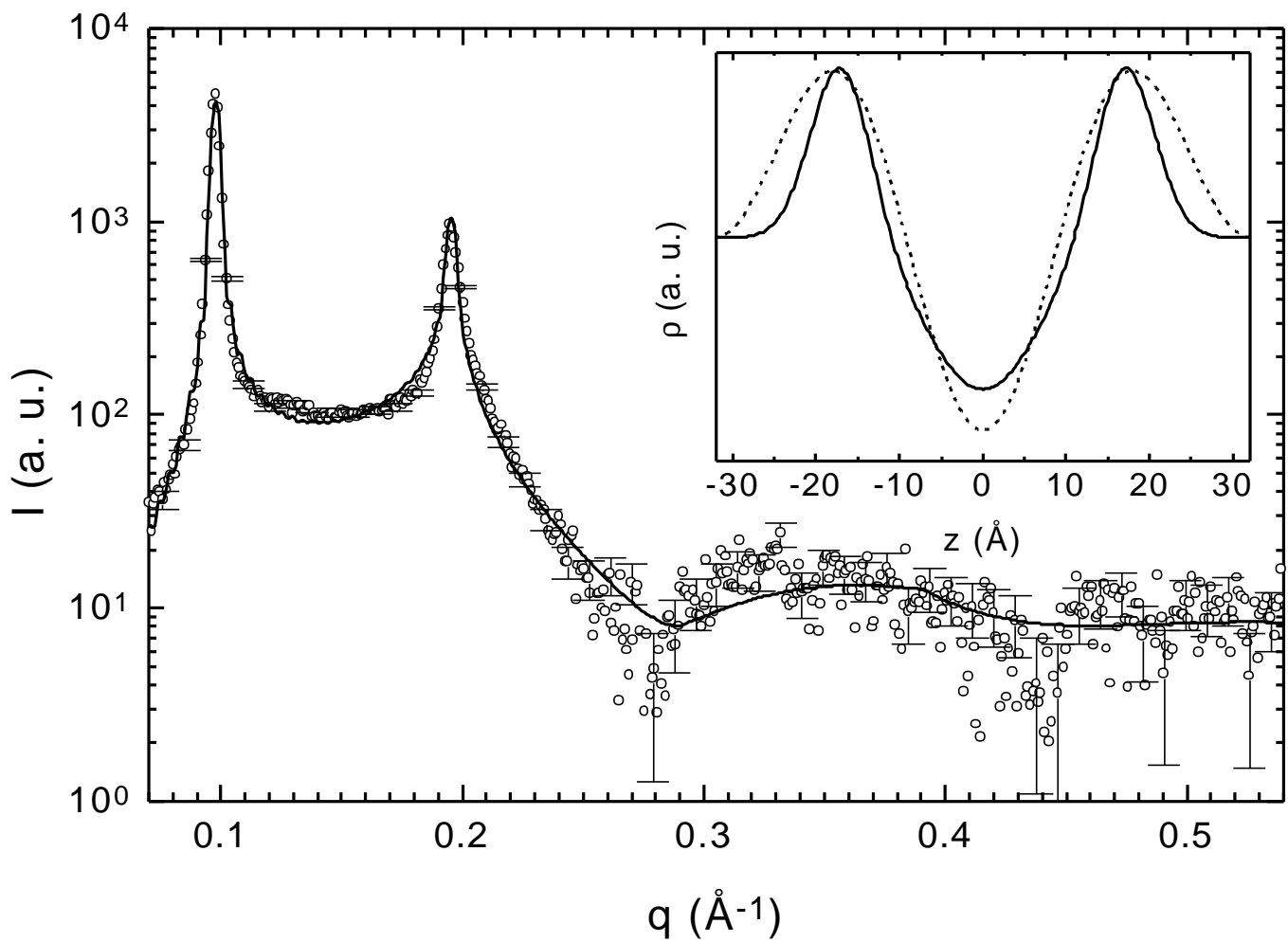

**FIG. 4** / Georg Pabst

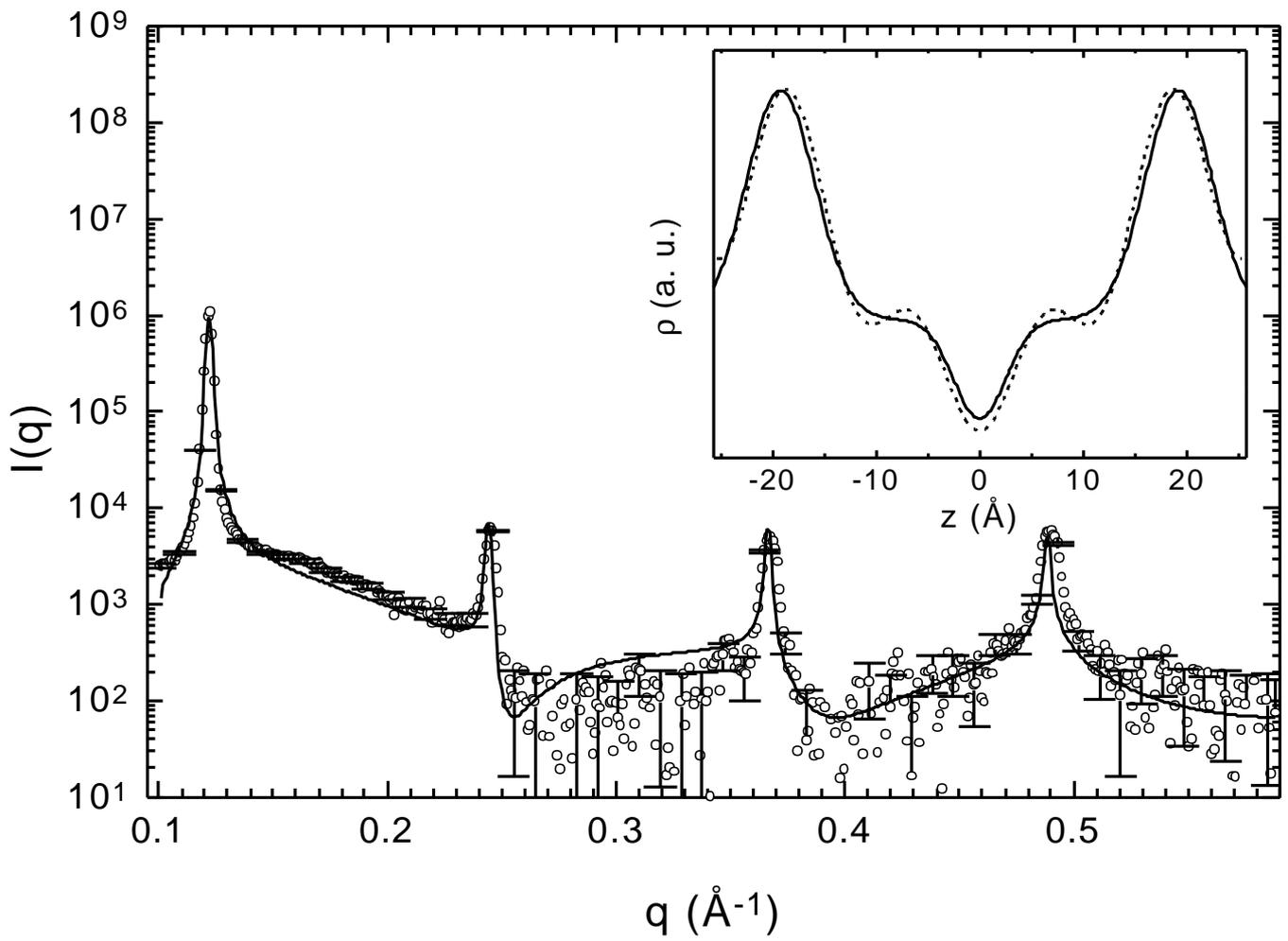

FIG. 5 / Georg Pabst

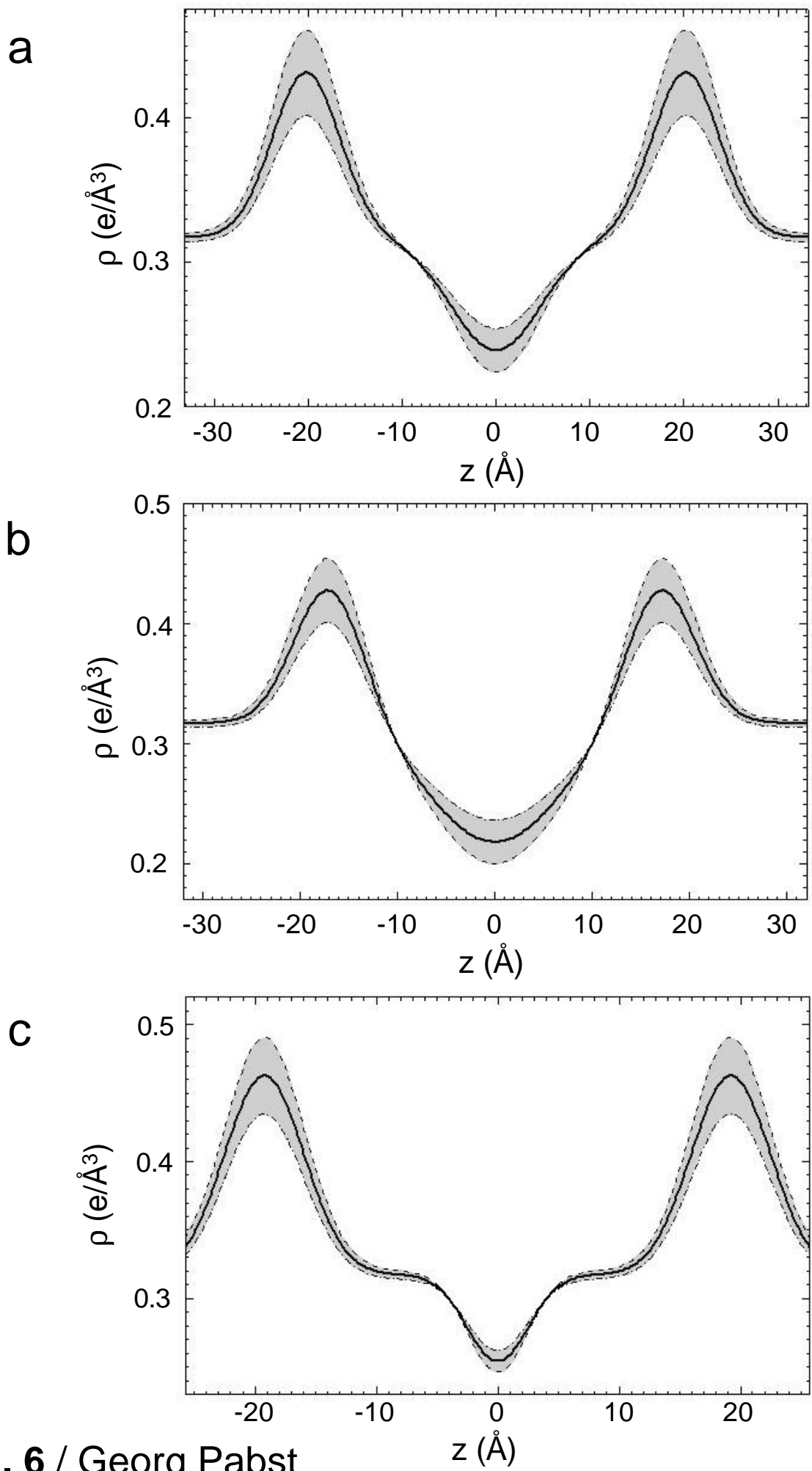

**FIG. 6** / Georg Pabst